\documentclass[twocolumn,secnumarabic,amssymb, amsmath, nobibnotes, showpacs, superscriptaddress, aps, prl, 10pt]{revtex4-1}
\usepackage{graphicx}
\usepackage{amsfonts} 
\usepackage{bm}
\usepackage[cm]{fullpage}

\usepackage{textgreek}

\newcommand{\fig}[1]{Fig.~\ref{#1}}

\begin{document}

\title{Optically Mediated Hybridization Between Two Mechanical Modes}

\author{A. B. Shkarin}
\email{alexey.shkarin@yale.edu}
\author{N. E. Flowers-Jacobs}
\author{S. W. Hoch}
\author{A. D. Kashkanova}
\affiliation{Department of Physics, Yale University, New Haven, CT  06520, USA}
\author{C. Deutsch}
\author{J. Reichel}
\affiliation{Laboratoire Kastler Brossel, ENS/UPMC-Paris 6/CNRS, F-75005 Paris, France}
\author{J. G. E. Harris}
\affiliation{Department of Physics, Yale University, New Haven, CT  06520, USA}
\affiliation{Department of Applied Physics, Yale University, New Haven, CT  06520, USA}
\date{\today}

\begin{abstract}
In this paper we study a system consisting of two nearly degenerate mechanical modes that couple to a single mode of an optical cavity.  We show that this coupling leads to nearly complete (99.5\%) hybridization of the two mechanical modes into a bright mode that experiences strong optomechanical interactions and a dark mode that experiences almost no optomechanical interactions.  We use this hybridization to transfer energy between the mechanical modes with 40\% efficiency.
\end{abstract}


\maketitle

  Optomechanical systems, in which electromagnetic resonators interact with mechanical resonators, offer a platform for studying a wide range of nonlinear and quantum effects. These systems have been studied in the context of quantum-limited detection of forces and displacements, the production of nonclassical states of light, synchronization and chaotic dynamics, and tests of quantum mechanics with massive degrees of freedom.\cite{AspelmeyerKippenbergMarquardtArXiv2013}
 
  Optomechanical systems are usually modeled as a single optical mode that is parametrically coupled to a single mechanical mode. This simple model accurately describes many experiments; however, real devices invariably consist of multiple optical and mechanical modes. The presence of multiple modes can provide important capabilities, including new types of optomechanical interactions, robust means for detecting quantum effects, and the ability to transfer quantum states between different systems.\cite{BraginskyScience1980,ThompsonNature2008,Hartmann2008,Genes2008,VerlotPRL2009,SankeyNatPhys2010,Ludwig2010,Regal2011,Heinrich2011, Seok2011,MeystreQuadratic2013, Xuereb2013}

  One important class of multimode optomechanical systems consists of devices in which a single optical mode couples to multiple mechanical modes. This situation arises naturally when an optomechanical device with well-separated optical resonances is driven by a single laser beam. Within the usual weak-coupling description of optomechanics the undriven optical modes are irrelevant, and only the driven mode needs to be considered.\cite{WilsonRae2007,Marquardt2007,CKLawPRA2011} Mechanical modes, on the other hand, cannot be ignored just because they are not driven. This is because any optical mode can be detuned (to some degree) by the displacement of any of the devices' mechanical modes. As a result the effective Hamiltonian for such a device will involve one optical mode coupled to many mechanical modes.

  In such a system, the motion of a given mechanical mode will modulate the intracavity optical field, which will in turn drive the other mechanical modes.  This can be thought of as an optically mediated coupling between the mechanical modes. This intermode coupling can be neglected for mechanical modes whose resonance frequencies are well separated. However, mechanical resonators with some degree of symmetry will have some nearly degenerate modes, and for these modes this coupling can be important.

  In this paper we demonstrate that the optomechanical coupling between one optical mode and two mechanical modes causes the mechanical modes to nearly fully (99.5\%) hybridize into bright and dark states.  We then transfer classical mechanical energy between the mechanical modes by modulating the hybridization in a classical analogy to Rabi oscillations.  The optomechanical hybridization of mechanical modes has been seen in a photonic double-nanobeam system\cite{Lin2010}, whispering gallery-mode resonators\cite{Lin2010, Zhang2012}, and nano-beams embedded in a microwave cavity\cite{Massel2012}.  However, these experiments did not use this hybridization to transfer energy.  Two of these devices would have a low transfer efficiency because of a relatively low mechanical quality factor\cite{Lin2010} or incomplete hybridization\cite{Massel2012}.  We estimate that the device in reference \cite{Zhang2012} could transfer energy with reasonable efficiency, but reference \cite{Zhang2012} focused on using the optical force to regeneratively oscillate and synchronize the two mechanical resonators.

\begin{figure}[t]
\includegraphics[width=.48\textwidth]{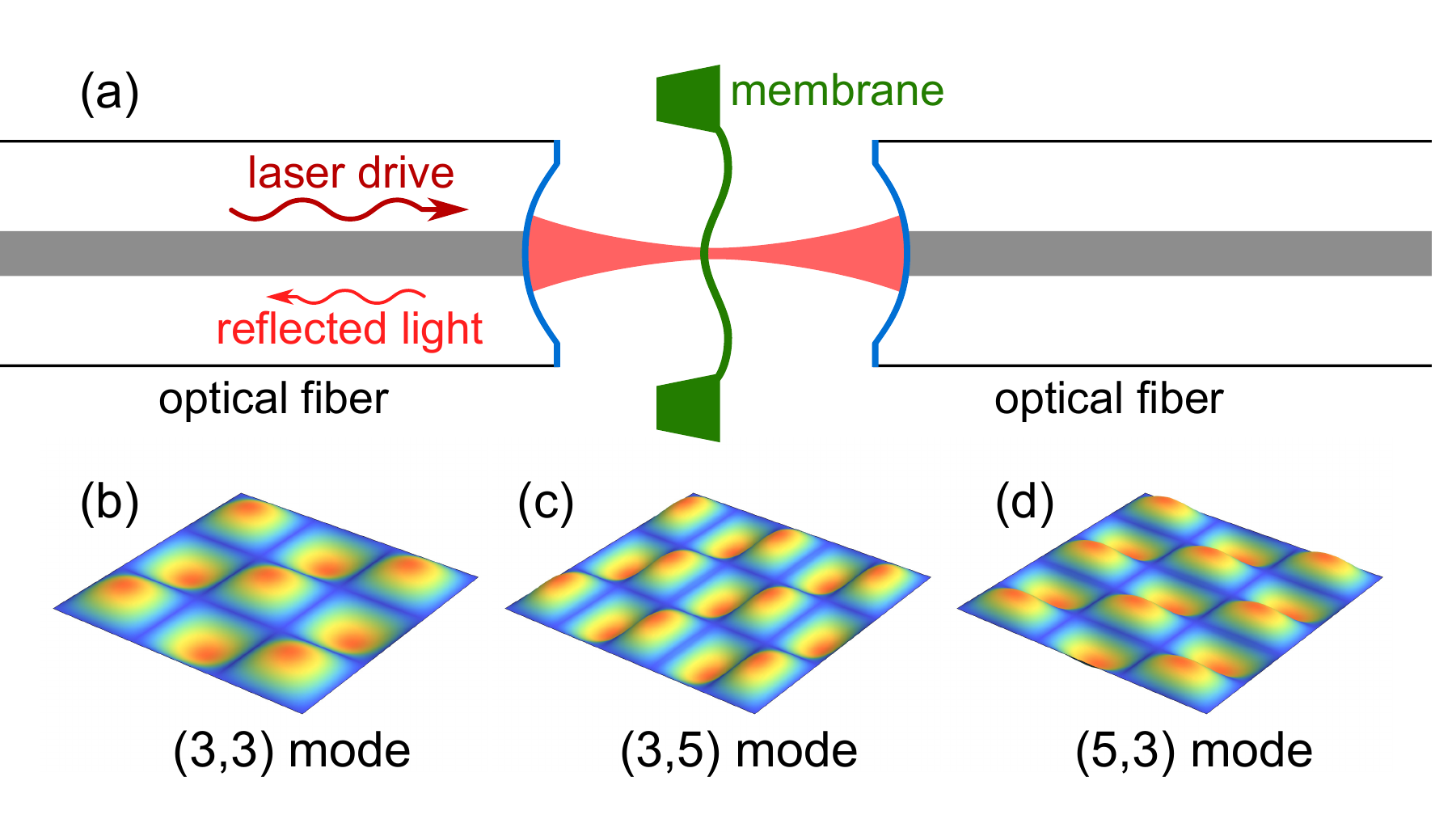}\\
\caption{(a): Experimental setup with a SiN membrane placed in a cavity formed between the mirrored ends of two fibers. (b)-(d): Schematic representation of the mode shapes of the three relevant membrane modes.}
\label{fig:setup_layout}
\end{figure}
   
  The device described here operates in the classical regime.  However, in the quantum regime (that is, when the mechanical modes are nearly in the ground state) the fact that the intrinsic mechanical damping rate is small and the intermode coupling is both conservative and strong (in contrast to previous work \cite{Lin2010, Massel2012, Zhang2012}) means that it would be well-suited for realizing proposals for entangling mechanical modes and creating non-classical mechanical states \cite{Hartmann2008, Genes2008, Ludwig2010, Seok2011}.  In addition, the long lifetime of the mechanically dark state could be used to store quantum information\cite{Lin2010, SafaviNaeini2011}.
  
  The device studied here is a ``membrane-in-the-middle'' optomechanical system composed of a SiN membrane placed in an optical fiber-cavity (\fig{fig:setup_layout}a)\cite{Hunger2010, Flowers-Jacobs2012}.  The 70~\textmu m long Fabry-Perot cavity is formed between the end faces of two 200~\textmu m diameter single-mode optical fibers.  Each fiber face has a concavity with a 300~\textmu m radius of curvature and a dielectric coating that is highly reflective at wavelength $\lambda=$1550~nm.  The resulting cavity has a finesse  $\mathtt{\lesssim}100,000$ depending upon the position of the membrane, corresponding to a cavity linewidth $\kappa/2\pi\mathtt{\gtrsim}$20 MHz.

  The SiN membrane is 250~\textmu m square and 100 nm thick. Because it is nearly square and under significant stress, the resonance frequencies of its higher-order modes are expected to be simply related to its fundamental resonance frequency $\omega_{\mathrm{(1,1)}}/2\pi=$1.7~MHz.  Labeling each mode by the number of anti-nodes along each axis $(j,k)$, as shown in \fig{fig:setup_layout}, the resonance frequencies are $\omega_{(j,k)}=\omega_{\mathrm{(1,1)}}\sqrt{j^2+k^2}/\sqrt{2}$.  We find that the measured $\omega_{(j,k)}$ follow this relationship to within 0.1\% for $j,k<6$, implying that each mode with $j\ne k$ has a (nearly) degenerate partner.
 
  As described in \cite{Flowers-Jacobs2012}, the membrane is positioned so that the frequency of the optical cavity varies linearly with the membrane position.  The cavity is locked to the laser at frequencies $\ll\omega_{\mathrm{(1,1)}}$ so the membrane's motion is imprinted on the reflected laser field, which is measured using a heterodyne technique.  We measure the power spectral density of the heterodyne signal near the membrane's resonance frequencies and fit this data to extract each mechanical mode's linewidth and resonance frequency.
  
  Before concentrating on the cavity-induced coupling between nearly degenerate mechanical modes, we characterize the optomechanical shift in the resonance frequency (``optical spring'') and linewidth (``optical damping'') of the non-degenerate $(3,3)$ mode. For this mode, $\omega_{\mathrm{(3,3)}}/2\pi=5.092$~MHz and the quality factor $Q_{\mathrm{(3,3)}}=500,000$.  The effective mass is $m = \rho V/4 = 5.4$~ng, which is the same for all of the membrane's modes.

\begin{figure}[t]
\centering
\includegraphics[width=.47\textwidth]{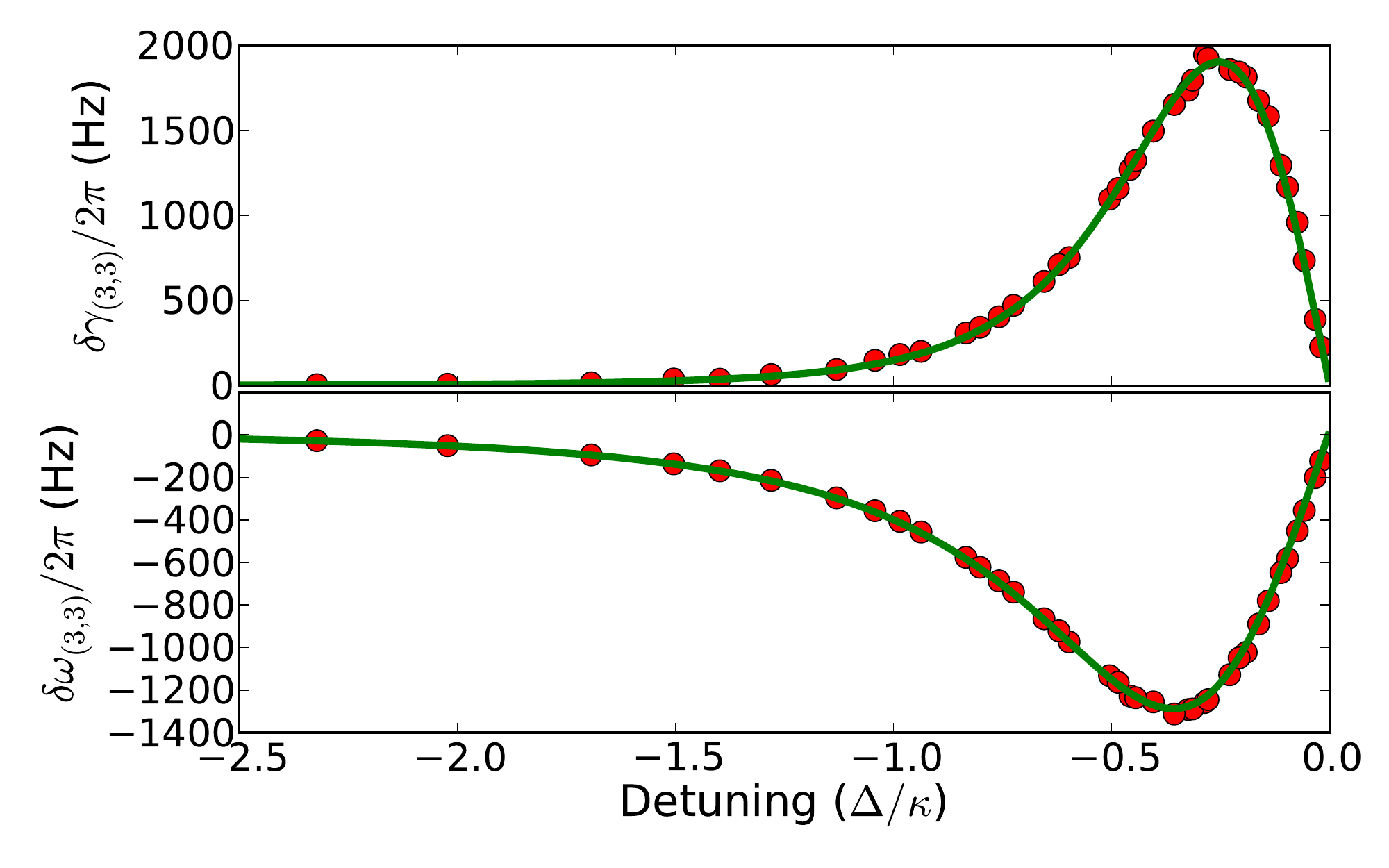}
\caption{Optomechanically-induced shift in mechanical linewidth (\textit{top}) and frequency (\textit{bottom}) of the (3,3) membrane mode as a function of detuning with theoretical fit (green line).  This data is taken with an incident power of 3~\textmu W and a cavity linewidth of 21 MHz.}
\label{fig:33_optomechanics}
\end{figure}

  The effects of the optomechanical coupling are revealed by varying the detuning $\Delta$ between the laser and the cavity.  In \fig{fig:33_optomechanics} we plot the shift in the mechanical linewidth $\delta \gamma_{\mathrm{(3,3)}}$ and the resonance frequency $\delta \omega_{\mathrm{(3,3)}}$ as a function of $\Delta$.  Since $\omega_{\mathrm{(3,3)}}\approx 0.2\kappa$ (the unresolved-sideband regime), $\delta \omega_{\mathrm{(3,3)}}$ and $\delta \gamma_{\mathrm{(3,3)}}$ are largest when $\Delta\approx-\kappa/3$.  We separately measure the incident power $P_{\mathrm{in}}=3$~\textmu W and relative input coupling  $\kappa_L=0.05\kappa$ and fit the data in \fig{fig:33_optomechanics} to theoretical predictions \cite{WilsonRae2007,Marquardt2007} using the single-photon optomechanical coupling $g_{(3,3)}$ and cavity linewidth $\kappa$ as fitting parameters. The result of this fit is shown in \fig{fig:33_optomechanics} (green line) and gives $\kappa/2\pi=21$~MHz, and $g_{\mathrm{(3,3)}}/2\pi=1050$~Hz, in agreement with independent measurements.
  
\begin{figure*}[t]
\centering
\includegraphics[width=1\textwidth]{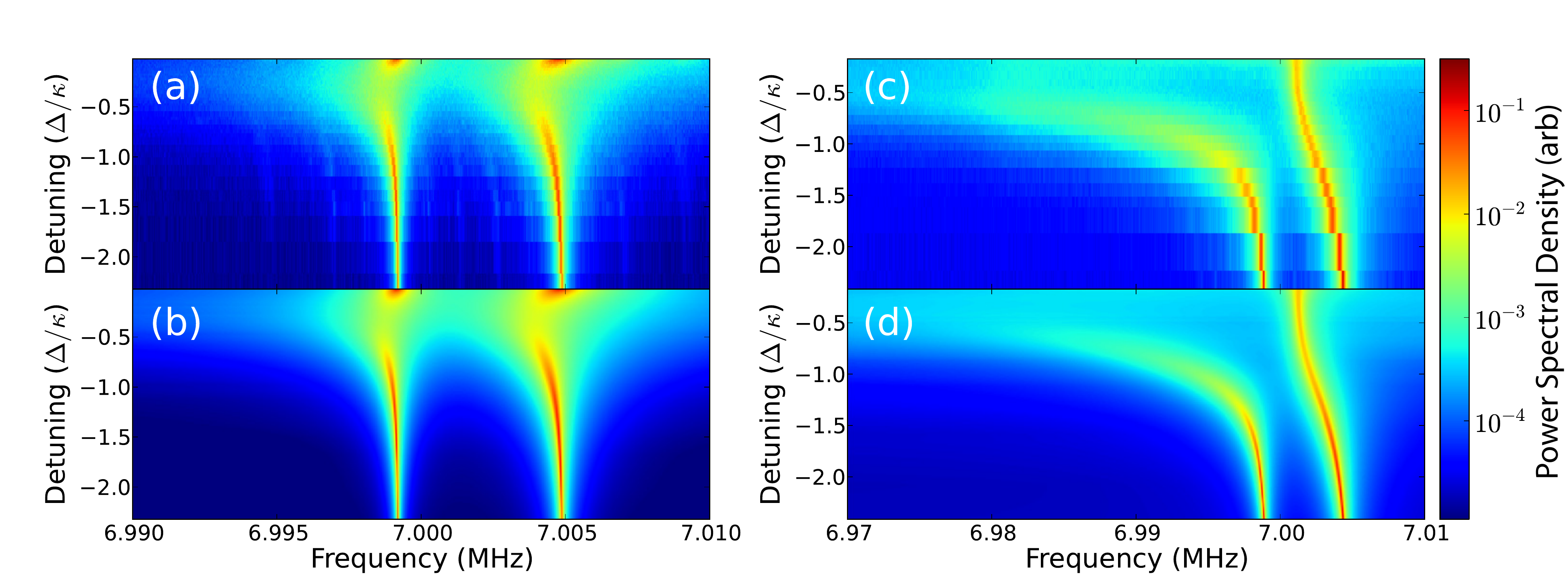}
\caption{Power spectral density of the heterodyne signal (a,c) and theoretical fits (b,d) as a function of measurement frequency (horizontal axis) and detuning between the incident laser and the cavity resonance (vertical axis). The data is presented for two incident laser powers: 3~\textmu W for (a,b) and 38~\textmu W for (c,d).   A direct comparison of the theory and data is shown in the supplemental material.}
\label{fig:35_PSDs}
\end{figure*}

  Now we focus on the effect of the optomechanical coupling on the nearly degenerate $(3,5)$ and $(5,3)$ mechanical modes.  For these modes $\omega_{\mathrm{(3,5)}}/2\pi=6.999$~MHz, $\omega_{\mathrm{(5,3)}}/2\pi =7.005$~MHz, $Q_{\mathrm{(3,5)}}=440,000$, $Q_{\mathrm{(5,3)}}=220,000$, $g_{\mathrm{(3,5)}}/2\pi=700$~Hz, and $g_{\mathrm{(5,3)}}/2\pi=950$~Hz.  In \fig{fig:35_PSDs}(a,c) we plot the measured power spectral density of the heterodyne signal as a function of $\Delta$ (y-axis) and the measurement frequency (x-axis) at two different incident powers ($P_{\mathrm{in}}=3$~\textmu W and $P_{\mathrm{in}}=38$~\textmu W).  The thermal motion of each mode is clearly visible in these power spectral densities.
  
  In order to qualitatively understand the data in \fig{fig:35_PSDs}(a,c) and make a comparison with theory, we consider a system of $N$ mechanical oscillators coupled to a single optical mode.  This analysis is presented in the supplemental material.  When $N=2$, as in our system, we can simplify the more general theory using a description based on bright and dark states.

  Specifically, we start with two intrinsic mechanical modes each with displacement $z_\mathrm{n}$, single-photon optomechanical coupling $g_\mathrm{n}$, intrinsic complex resonance frequency $\xi_\mathrm{n}=\omega_\mathrm{n}-i\gamma_\mathrm{n}/2$, and intrinsic mechanical susceptibility $\chi_\mathrm{n}[\omega]^{-1}=i\xi_\mathrm{n}-i\omega$ (where $n=1,2$).  We then define a dark state displacement $z_\mathrm{d}=v z_\mathrm{1}-u z_\mathrm{2}$ which is a linear combination of the original, intrinsic mode displacements with weights $u,v=g_{\mathrm{1,2}}/\sqrt{g_\mathrm{1}^2+g_\mathrm{2}^2}$.  The ``dark'' label is used because $z_\mathrm{d}$ is not coupled to the cavity (that is, the single-photon optomechanical coupling $g_\mathrm{d}=0$).
  
  On the other hand, the single-photon optomechanical coupling of the bright mode, with modal displacement $z_\mathrm{b}=u z_\mathrm{1}+v z_\mathrm{2}$, is larger than that of the original modes $g_\mathrm{b}=\sqrt{g_\mathrm{1}^2+g_\mathrm{2}^2}$.  The new modes $z_\mathrm{b}$ and $z_\mathrm{d}$ have intrinsic complex resonance frequencies $\xi_\mathrm{b,d}=u^2\xi_\mathrm{1,2}+v^2\xi_\mathrm{2,1}$ and are generally not normal modes of the system; the effective coupling between them is $g_{\mathrm{bd}}=uv(\xi_\mathrm{1}-\xi_\mathrm{2})$.
  
  Using these expressions, the displacement spectra of $z_\mathrm{b}$ and $z_\mathrm{d}$ in response to the thermal Langevin forces $\eta_\mathrm{b}$ and $\eta_\mathrm{d}$ are
\begin{eqnarray}
\label{eq:bright_mode_spectrum}
(\chi_\mathrm{b}^{-1}[\omega]+i\Sigma_\mathrm{bb}[\omega])z_\mathrm{b}[\omega]&=&-ig_{\mathrm{bd}}z_\mathrm{d}[\omega]+\sqrt{\gamma_\mathrm{b}}\eta_\mathrm{b}[\omega]\\
\label{eq:dark_mode_spectrum}
\chi_\mathrm{d}^{-1}[\omega]z_\mathrm{d}[\omega]&=&-ig_{\mathrm{bd}}z_\mathrm{b}[\omega]+\sqrt{\gamma_\mathrm{d}}\eta_\mathrm{d}[\omega].
\end{eqnarray}
The only term in these expressions that depends on the optical drive is the ``self-energy''  $\Sigma_\mathrm{bb}[\omega]$, which determines the optical spring $\delta\omega_{\mathrm{b}}=\mathrm{Re}(\Sigma_{\mathrm{bb}}[\omega_\mathrm{b}])$ and damping $\delta\gamma_{\mathrm{b}}=-2\mathrm{Im}(\Sigma_{\mathrm{bb}}[\omega_\mathrm{b}])$ of the bright mode.

  We use this model to fit the data in \fig{fig:35_PSDs}(a,c) and plot the resulting theoretical curves in \fig{fig:35_PSDs}(b,d) (see supplemental material for a direct comparison of theory and data). The system parameters $\kappa$, $\Delta$, and $P_{\mathrm{in}}$ are determined from simultaneous measurements of the (3,3) mode (as in \fig{fig:33_optomechanics}).  We then use a least-squared fit to the data in \fig{fig:35_PSDs}(a,c) to determine the remaining parameters: $g_{1,2}$, $\omega_{1,2}$, and $\gamma_{1,2}$ (where the subscripts 1 and 2 now label the modes (3,5) and (5,3)).

   This model also provides a qualitative interpretation of the data.  In order to significantly hybridize the intrinsic modes into bright and dark modes, the optical spring $\delta\omega_{\mathrm{b}}$ needs to be large enough that $|\omega_\mathrm{b}+\delta\omega_{\mathrm{b}}-\omega_\mathrm{d}|\gg|g_{\mathrm{bd}}|$ or, in this case, $-\delta\omega_{\mathrm{b}}/2\pi\gg1$~kHz.  At low $P_{\mathrm{in}}$ (\fig{fig:35_PSDs}(a,b)) or at high $P_{\mathrm{in}}$ and large detunings (near the bottom of \fig{fig:35_PSDs}(c,d)), the optical spring is relatively small and this condition is not satisfied.  The intrinsic modes do not significantly hybridize and instead independently exhibit essentially the same behavior as shown in \fig{fig:33_optomechanics} for the nondegenerate $(3,3)$ mode.

\begin{figure*}[ht!]
\centering
\includegraphics[width=1.\textwidth]{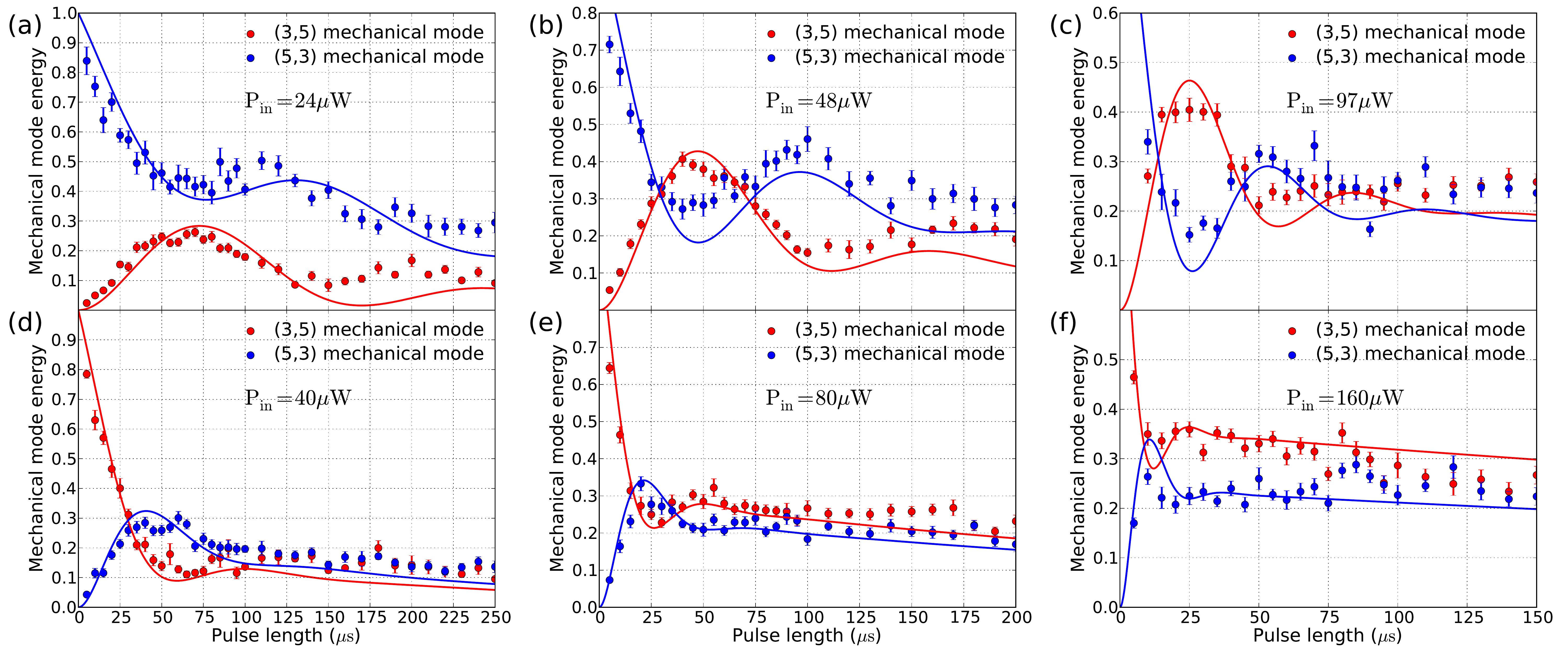}
\caption{The energy in each mechanical mode immediately after a hybridization pulse, plotted as a function of the pulse duration. The energy is normalized to the energy in the driven mode just before the hybridization pulse. In plots (a,b,c) the system is initialized by driving the (3,5) mode and shows the transfer of energy to the (5,3) mode.  Plots (d,e,f) show the transfer of energy in the opposite direction. Solid lines are the fits described in the text, and the error bars indicate statistical uncertainties.}
\label{fig:energy_transfer}
\end{figure*}

  On the other hand, in \fig{fig:35_PSDs}(c,d) at detunings $\Delta \gtrsim -1.5\kappa$ the optical spring is large enough that the intrinsic modes begin to hybridize into bright and dark modes.  When the detuning $\Delta \gtrsim -0.75\kappa$ the lower-frequency state is almost entirely bright and exhibits large optical spring and optical damping, while the higher frequency state is almost entirely dark (based on the fit parameters from \fig{fig:35_PSDs} the hybridization is 99.5\%).  In this regime the coupling $g_{\mathrm{bd}}$ leads to only two noticeable effects. First, it makes the effective dark mode linewidth larger than the intrinsic linewidth of $\gamma_\mathrm{d}/2\pi=20$~Hz. Second, it allows the dark mode to be visible in the reflected light spectrum; otherwise this mode would be completely uncoupled from the cavity field.

  The high mechanical quality factors and purely optomechanical coupling of the membrane modes make it possible to observe this hybridization in the time domain.  As shown in \fig{fig:energy_transfer}, modulating the optical drive results in the transfer of mechanical energy between the two intrinsic mechanical modes.

  This measurement starts by using a piezo to drive either the (3,5) or (5,3) mechanical mode and locking the cavity to a weak laser beam with detuning $\Delta_{\mathrm{weak}}=-0.7\kappa$ for \fig{fig:energy_transfer}(a,b,c) and $\Delta_{\mathrm{weak}}=-0.4\kappa$ for \fig{fig:energy_transfer}(d,e,f). For this measurement, the cavity linewidth $\kappa/2\pi=40$~MHz.  The weak laser beam is primarily used to measure the mechanical displacement, though its dynamical backaction does increase the mechanical linewidths by a factor $\sim 2$.  The piezo drive is turned off and a strong laser beam at detuning $\Delta_{\mathrm{weak}}+\kappa/8$ and power $P_{\mathrm{in}}$ is turned on for a time $\tau$. This pulse hybridizes the mechanical modes.  After this pulse, the weak laser beam is used to determine the energy in each of the intrinsic mechanical modes.  This measurement is facilitated by the separation in time scales between the 10~ms lifetimes of the intrinsic mechanical modes, the 100~\textmu sec period of the hybridization oscillations, the 140~ns period of mechanical oscillations, and the 10~ns lifetime of the optical cavity.

  In \fig{fig:energy_transfer} we plot the ratio of the final energy in each intrinsic mechanical mode after the pulse to the total initial energy as a function of $\tau$ and at different $P_{\mathrm{in}}$. In \fig{fig:energy_transfer}(a,b,c) the system is initialized by driving the (5,3) mode, while in \fig{fig:energy_transfer}(d,e,f) it is initialized by driving the (3,5) mode.  The theory curves in \fig{fig:energy_transfer} are derived from the solution to a set of differential equations describing the motion $z_\mathrm{1}$ and $z_\mathrm{2}$ of two linearly coupled harmonic oscillators.  The coupling and oscillator parameters are taken from the ``self-energy'' matrix $\Sigma[\omega]$ (see supplemental material) and depend on the strength and detuning of the ``strong'' laser pulse.

  In \fig{fig:energy_transfer}, some of the parameters for the theory curves are chosen manually to match the data.  The values of $g_{1,2}$, $\omega_{1,2}$, and $\gamma_{1,2}$ are determined by fitting data similar to \fig{fig:35_PSDs}.  The cavity linewidth $\kappa$ is measured independently.  A single value of $\Delta$ is chosen to fit the data in the three upper plots. $P_{\mathrm{in}}$ is chosen to fit the data in \fig{fig:energy_transfer}(a), and then increased by a factor of two in \fig{fig:energy_transfer}(b), and another factor of two in \fig{fig:energy_transfer}(c), in accordance with the experimental procedure. The same approach was used to choose different values of $\Delta$ and $P_{\mathrm{in}}$ for the lower three plots.  Finally, we apply a scaling factor of $1.3$ to the initial energy in the driven mode to correct for the nonlinearity of the detector.  This manual choice of five parameters completely determines the theory curves in \fig{fig:energy_transfer}.
  
  The pulse power used in \fig{fig:energy_transfer} is sufficient to hybridize the system, resulting in Rabi-like oscillations between the intrinsic (3,5) and (5,3) eigenmodes.  We can gain a more qualitative understanding of the data in \fig{fig:energy_transfer} by considering the hybridization of the original modes into bright and dark modes.  The oscillation frequency increases with $P_{\mathrm{in}}$ since the frequency splitting between the dark and bright modes is increased.  The oscillations are suppressed on a time scale given by the optomechanically-dominated damping rate of the bright mode, which also increases with increasing $P_{\mathrm{in}}$.  After the bright mode decays, the ratio of the energy in the two modes is constant and given by the fractional contribution of each intrinsic mode to the dark mode.  The total energy continues to decrease as the dark mode decays.

  By optimizing the pulse power and length, we are able to transfer energy between the two intrinsic modes with an efficiency of 40\% (e.g., in \fig{fig:energy_transfer}(b,c)).  This transfer efficiency is limited by the optomechanical damping of the bright mode $\delta\gamma_b$.  Since $\delta\gamma_b$ is comparable to the coupling rate between the mechanical modes, significant energy is lost to the optical field during the energy transfer.  The transfer efficiency can be increased by increasing the ratio of the optical spring to the optical damping $\delta\omega_b/\delta\gamma_b$ by, for example, operating in either the resolved $\kappa \ll \omega_m$ or unresolved $\kappa \gg \omega_m$ sideband limit \cite{Seok2011}.

  The main barrier between the present setup and operation in the quantum regime is the 300~K temperature of the environment.  To consider the performance of this system in a cryogenic environment, we note that if it was cooled to 100~mK, it would be possible to laser-cool both of the membrane modes to a mean energy of less than one phonon.\cite{WilsonRae2007,Marquardt2007} Assuming that $Q_{\mathrm{(3,5)}}$ and $Q_{\mathrm{(5,3)}}$ increase to $5\times10^6$ at cryogenic temperatures\cite{Jayich2012,Purdy2012}, then the thermal and optomechanically-induced decoherence rates become comparable to the coupling strength between the two mechanical modes.  With these assumptions, we estimate the quantum state transfer fidelity to be $~10\%$ (see supplemental material).  We note that the device described here is well-suited to cryogenic operation.  For example, SiN membranes have been used in a number of cryogenic optomechanical experiments\cite{Jayich2012,Purdy2012} and we have shown that fiber-cavities can operate at 4~K (see supplemental material).

  We thank H. Seok and P. Meystre for helpful discussions related to quantum state transfer.  This work has been supported by the DARPA/MTO ORCHID program through a grant from AFOSR.

\newpage

\onecolumngrid

\begin{center}

\large{\textbf{Supplemental material:}\\
\textbf{Optically Mediated Hybridization Between Two Mechanical Modes}}

\vspace{0.2in}

\normalsize{A. B. Shkarin,$^{1,*}$ N. E. Flowers-Jacobs,$^{1}$ S. W. Hoch,$^{1}$ A. D. Kashkanova,$^{1}$ C. Deutsch,$^{2}$ J. Reichel,$^{2}$ J. G. E. Harris$^{1,3}$}\\
\vspace{0.05in}
\small{$^1$ \textit{Department of Physics, Yale University, New Haven, CT 06520, USA}\\
$^2$ \textit{Laboratoire Kastler Brossel, ENS/UPMC-Paris 6/CNRS, F-75005 Paris, France}\\
$^3$ \textit{Department of Applied Physics, Yale University, New Haven, CT 06520, USA}\\}

\end{center}

\twocolumngrid

\begin{figure*}[!]
\centering
\includegraphics[width=1\textwidth]{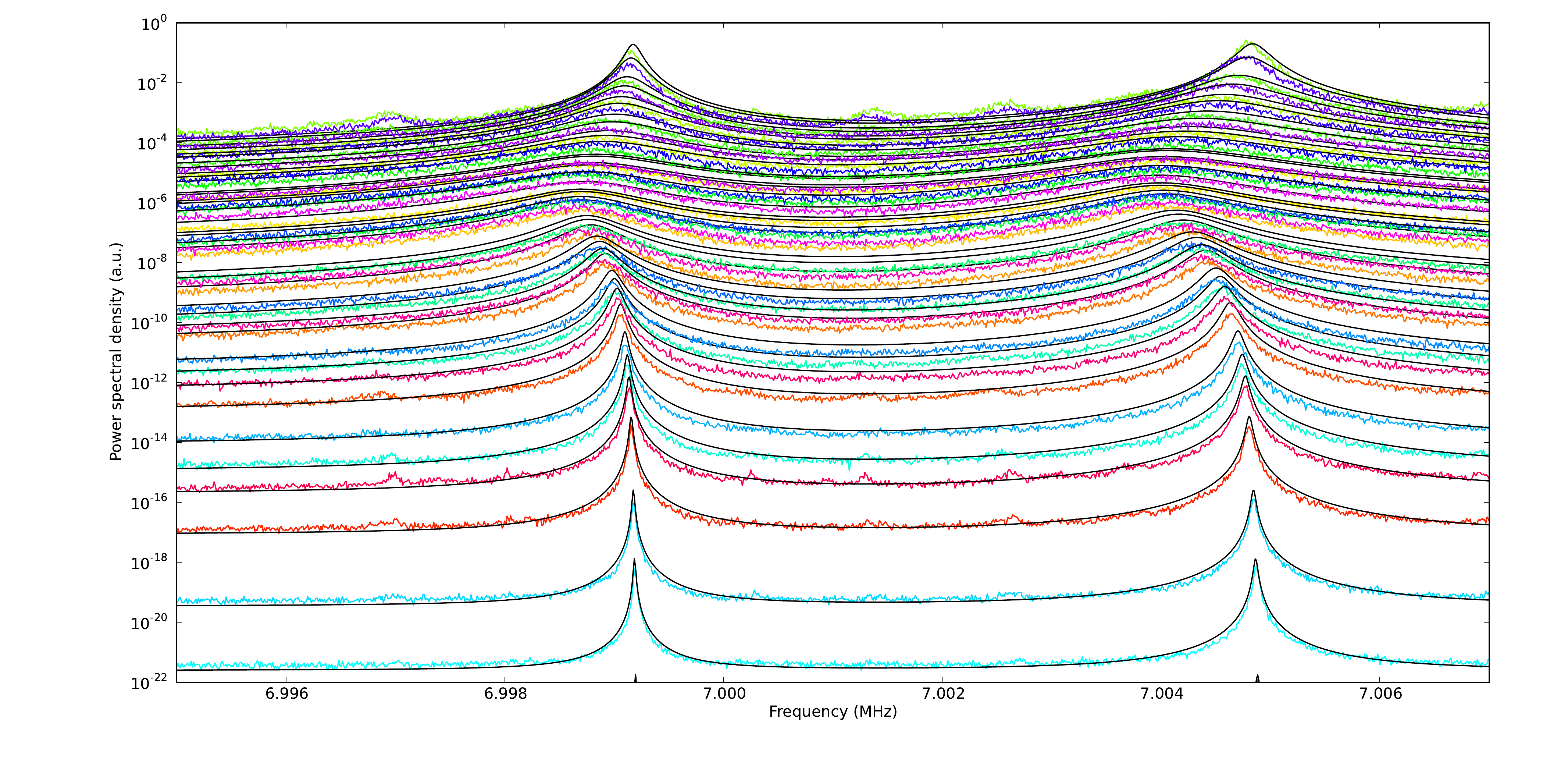}
\includegraphics[width=1\textwidth]{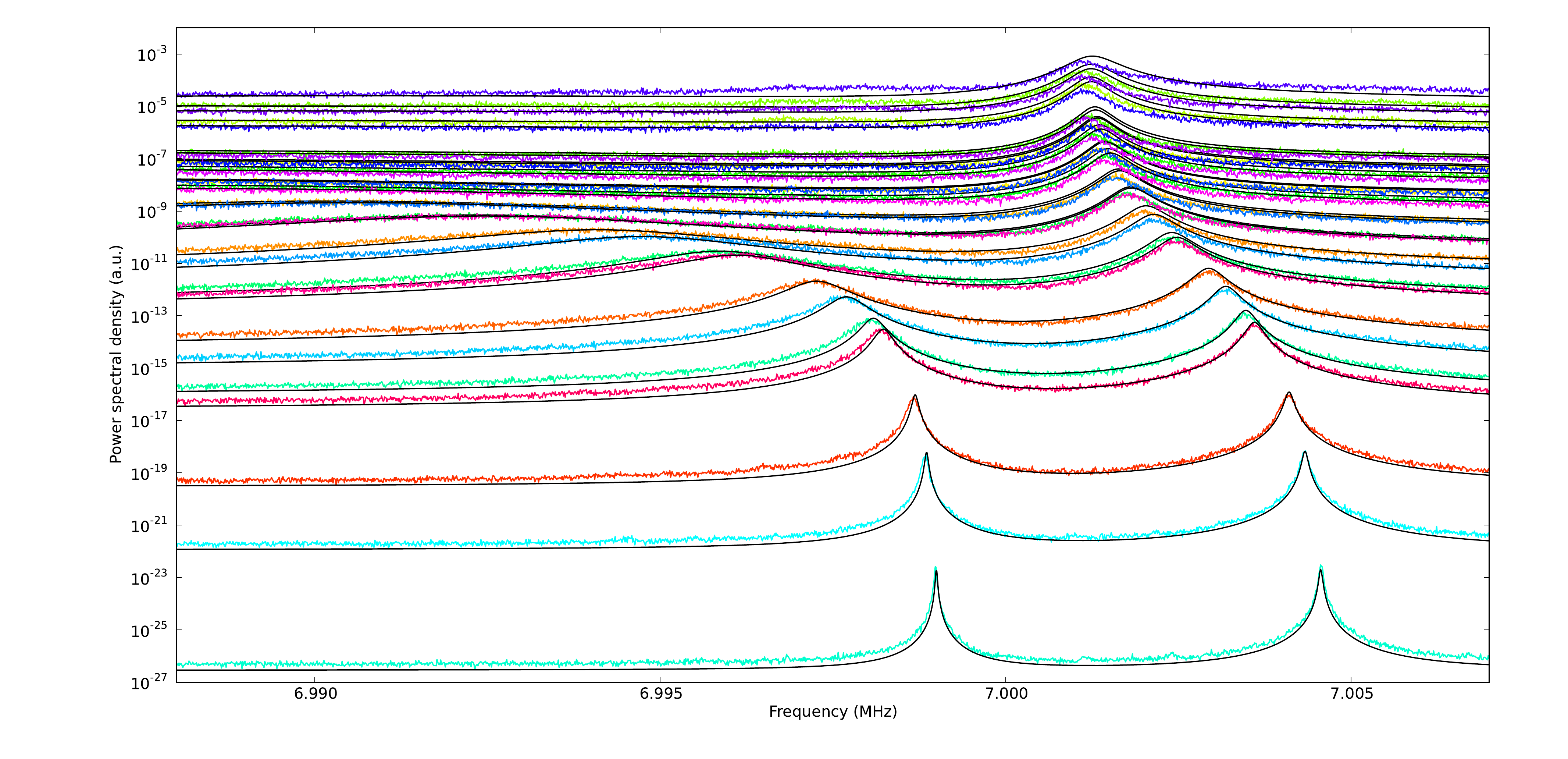}
\caption{Power spectral densities of reflected light showing the thermal motion of two mechanical modes.  Each data set is taken at a different detuning of the incident laser beam from the cavity resonance with either 3~\textmu W incident power (top) or 38~\textmu W incident power (bottom). For clarity, the experimental data (color lines) and fit to theory (black lines) at each detuning are offset along the y-axis by 1 decade per MHz of detuning (that is, 1 decade per $0.05\kappa$ of detuning).  The same experimental data and fits are presented as density plots in Fig. 3 of the main article; these supplemental plots are intended to facilitate a more detailed comparison of the measured power-spectral density and theoretical fits.  The theory and fitting procedure are discussed in the main article.}
\end{figure*}

\Large{\bf{1.  Theoretical derivation and fits to data}} \\ \\
\normalsize
  We consider a system with several mechanical modes, all of which couple to a single optical mode. The optical mode's linewidth is $\kappa=\kappa_\mathrm{L}+\kappa_\mathrm{M}$, where $\kappa_\mathrm{L}$ is the cavity loss rate due to coupling through the input mirror and $\kappa_\mathrm{M}$ includes all other losses. The cavity is driven by a laser with amplitude $a_{\mathrm{in}}$ which is detuned by $\Delta$ from the cavity resonant frequency $\omega_\mathrm{c}$. There are $N$ mechanical oscillators coupled to the cavity mode and the $n{\mathrm{th}}$ mechanical oscillator has a single-photon optomechanical coupling $g_\mathrm{n}$, intrinsic linewidth $\gamma_\mathrm{n}$, and resonant frequency $\omega_\mathrm{n}$. Since our room-temperature experiment is firmly in the classical regime $\hbar\omega_\mathrm{n}\ll k_\mathrm{B} T$, we use a classical analysis.  We also disregard all noise on the optical drive.  Under these assumptions, the system's linearized equations of motion are
\begin{eqnarray}
\label{eq:mm_linearized_motion_eqn_d}
\dot{d}&=&-(\frac{\kappa}2 -i\Delta)d-i\sum_{n}\alpha_\mathrm{n} z_\mathrm{n}\\
\label{eq:mm_linearized_motion_eqn_c}
\dot{c}_\mathrm{n}&=&-(\frac{\gamma_\mathrm{n}}2 +i\omega_\mathrm{n})c_\mathrm{n}-i(\alpha_\mathrm{n}^*d+\alpha_\mathrm{n} d^*)+\sqrt{\gamma_\mathrm{n}}\eta_\mathrm{n},
\end{eqnarray}
where $d=a-\bar{a}$ are the small fluctuations of the optical mode's amplitude around its mean value $\bar{a}=\frac{\sqrt{\kappa_\mathrm{L}}a_{\mathrm{in}}}{\kappa/2-i\Delta}$, $z_\mathrm{n}=c_\mathrm{n}^*+c_\mathrm{n}$ is the position of the $n{\mathrm{th}}$ mechanical oscillator, $c_n$ is the quadrature amplitude of the mode's displacement, and $\alpha_\mathrm{n}=\bar{a} g_\mathrm{n}$.  Each mode's thermal fluctuations are generated by a thermal Langevin force $\eta_\mathrm{n}(t)$ where $\langle\eta_\mathrm{n}^*(t)\eta_\mathrm{m}(t')\rangle=\delta_{\mathrm{nm}}\delta(t-t') k_\mathrm{B} T/\hbar\omega_\mathrm{n}$.

  From this model, we can recover the standard dynamical backaction for a single mechanical mode coupled to a cavity [12,13] by taking $N=1$.  To do this, we find the Fourier transform of the displacement $z_\mathrm{1}[\omega]$ arising from the thermal Langevin force by inserting the expression for $d$ from (\ref{eq:mm_linearized_motion_eqn_d}) into (\ref{eq:mm_linearized_motion_eqn_c}) and taking the limit $\omega_\mathrm{1}\gg \gamma_\mathrm{1}$:
\begin{equation}
\label{eq:z_spectrum}
(\chi_{\mathrm{1}}^{-1}[\omega]+i\Sigma_{\mathrm{11}}[\omega])z[\omega]=\sqrt{\gamma_\mathrm{1}}\eta_\mathrm{1}[\omega]
\end{equation}
where the intrinsic mechanical susceptibility $\chi_{\mathrm{1}}[\omega]^{-1}=\gamma_\mathrm{1}/2-i(\omega-\omega_\mathrm{1})$ is modified by an optically-induced ``self-energy'' term $\Sigma_{\mathrm{11}}[\omega]=-i|\alpha_\mathrm{1}|^2(\chi_\mathrm{c}[\omega]-\chi_\mathrm{c}^*[-\omega])$ which is a function of the cavity susceptibility $\chi_\mathrm{c}[\omega]=(\kappa/2-i(\omega+\Delta))^{-1}$.  This self-energy represents the optomechanical contribution to the mechanical resonance frequency $\delta\omega_{\mathrm{1}}=\mathrm{Re}(\Sigma_{\mathrm{11}}[\omega_\mathrm{1}])$ and damping $\delta\gamma_{\mathrm{1}}=-2\mathrm{Im}(\Sigma_{\mathrm{11}}[\omega_\mathrm{1}])$.  The $(3,3)$ mode data in Fig. 2 of the main article is fit to these expressions for $\delta\omega_{\mathrm{1}}$ and $\delta\gamma_{\mathrm{1}}$, which are consistent with more complete derivations of single-mode dynamical backaction [12,13].

  Extending this approach to the full system with $N$ oscillators coupled to a single cavity, the analog of (\ref{eq:z_spectrum}) is
\begin{equation}
\label{eq:mm_z_spectra}
(\chi_{\mathrm{n}}^{-1}[\omega]+i\Sigma_{\mathrm{nn}}[\omega])z_\mathrm{n}[\omega]=-i\sum_{m\neq n}\Sigma_{\mathrm{nm}}[\omega]z_\mathrm{m}[\omega]+\sqrt{\gamma_\mathrm{n}}\eta_\mathrm{n}[\omega]
\end{equation}
where $\Sigma[\omega]$ is now a matrix with elements  $\Sigma_{\mathrm{nm}}[\omega]=-i|\alpha_\mathrm{n}\alpha_\mathrm{m}|(\chi_\mathrm{c}[\omega]-\chi_\mathrm{c}^*[-\omega])$.  In solving for $\Sigma[\omega]$ we continue to assume that $\omega_\mathrm{n}\gg \gamma_\mathrm{n}$ and that $g_\mathrm{n}$ is a real positive number for all $n$ so that $\alpha_\mathrm{n}\alpha_\mathrm{m}^*=|\alpha_\mathrm{n}\alpha_\mathrm{m}|=\alpha_\mathrm{n}^*\alpha_\mathrm{m}$. This set of equations describes a system of coupled harmonic oscillators where all of the oscillator parameters and coupling rates depend on the intra-cavity field.

  Equation (\ref{eq:mm_z_spectra}) describes the behavior of the $N$ mechanical oscillator system, but we can better understand the behavior of our system (in which $N=2$) by using a bright and dark state description.  Specifically, we define a dark state displacement 
\begin{eqnarray}
\label{eq:dark_mode_def}
z_\mathrm{d}=v z_\mathrm{1}-u z_\mathrm{2}
\end{eqnarray}
which is a linear combination of the original, intrinsic mode displacements $z_\mathrm{1}$ and $z_\mathrm{2}$ with weights $u=g_{\mathrm{1}}/\sqrt{g_\mathrm{1}^2+g_\mathrm{2}^2}$ and $v=g_{\mathrm{2}}/\sqrt{g_\mathrm{1}^2+g_\mathrm{2}^2}$.  The ``dark'' label is used because $z_\mathrm{d}$ is not coupled to the cavity (that is, the single-photon optomechanical coupling $g_\mathrm{d}=0$).  On the other hand, the single-photon optomechanical coupling of the bright mode, with modal displacement
\begin{eqnarray}
\label{eq:bright_mode_def}
z_\mathrm{b}=u z_\mathrm{1}+v z_\mathrm{2},
\end{eqnarray}
is larger than that of the original modes $g_\mathrm{b}=\sqrt{g_\mathrm{1}^2+g_\mathrm{2}^2}$.

  The new modes $z_\mathrm{b}$ and $z_\mathrm{d}$ are generally not normal modes of the system, so the final ingredient in the bright/dark state description is the effective coupling between the modes $g_{\mathrm{bd}}=uv(\xi_\mathrm{1}-\xi_\mathrm{2})$ where $\xi_n=\omega_n-i\gamma_n/2$ is the complex resonance frequency of the $n{\mathrm{th}}$ mode (where $n = 1,2$).  The bright and dark states have $\xi_\mathrm{b}=u^2\xi_\mathrm{1}+v^2\xi_\mathrm{2}$ and $\xi_\mathrm{d}=u^2\xi_\mathrm{2}+v^2\xi_\mathrm{1}$.  Using these expressions, the displacement spectra $z_\mathrm{b}[\omega]$ and $z_\mathrm{d}[\omega]$ are
\begin{eqnarray}
\label{eq:bright_mode_spectrum}
(\chi_\mathrm{b}^{-1}[\omega]+i\Sigma_\mathrm{bb}[\omega])z_\mathrm{b}[\omega]&=&-ig_{\mathrm{bd}}z_\mathrm{d}[\omega]+\sqrt{\gamma_\mathrm{b}}\eta_\mathrm{b}[\omega]\\
\label{eq:dark_mode_spectrum}
\chi_\mathrm{d}^{-1}[\omega]z_\mathrm{d}[\omega]&=&-ig_{\mathrm{bd}}z_\mathrm{b}[\omega]+\sqrt{\gamma_\mathrm{d}}\eta_\mathrm{d}[\omega].
\end{eqnarray}
Note that the only term in these expressions that depends on the optical drive is 
\begin{eqnarray}
\Sigma_\mathrm{bb}[\omega]=-i|\bar{a}|^2g_b^2(\chi_c[\omega]-\chi_c^*[-\omega]),
\end{eqnarray}
which determines the optical spring and damping of the bright mode.
  We use this model to fit the data in Fig. 3(a,c) of the main article and plot the resulting theoretical predictions in Fig. 3(b,d) of the main article.  The same data and theory is plotted in a different format in Fig. 1 of the supplemental material.
  
\clearpage

  \Large{\bf{2.  Transfer efficiency and moving towards the quantum regime }} \\ \\
\normalsize
  In this experiment, the efficiency of energy transfer between the two intrinsic mechanical modes is predominately limited by the optomechanical damping of the bright mode $\delta\gamma_b$.  This damping rate is comparable to the coupling rate between the mechanical modes (the real part of the off-diagonal self-energy matrix element $\mathrm{Re}(\Sigma_{12})$, see (\ref{eq:mm_z_spectra})).  Since the coupling rate is also approximately equal to the rate at which mechanical energy oscillates between the two intrinsic modes, the bright mode loses significant energy to the environment during the energy transfer.
  
  This problem can be alleviated by reducing the magnitude of the optical damping $\delta\gamma_b$ relative to the optical spring $\delta\omega_b$.  This occurs naturally when operating deep in either the resolved $\kappa \ll \omega_m$ or unresolved $\kappa \gg \omega_m$ sideband limit and appropriately choosing the detuning and power of the coupling beam, as described in \cite{Seok2011}.
  
  A more fundamental restriction on the energy transfer efficiency is the intrinsic mechanical damping.  Even when the optical damping is small, the transfer efficiency will be $<10\%$ if the sum of the two intrinsic mechanical linewidths is comparable to the mechanical coupling strength.  Another requirement for complete energy transfer is that the two optically-shifted mechanical modes must have the same frequency and damping ($\omega_1+\delta\omega_1=\omega_2+\delta\omega_2$ and $\gamma_1+\delta\gamma_1=\gamma_2+\delta\gamma_2$). If this condition is not approximately satisfied, then the energy transfer will not be efficient even in the absence of any dissipation in the system; this is similar to Rabi-flopping with an off-resonant drive.

  When quantum state transfer is considered, the system requirements become much stricter.  Our room-temperature system is unable to access the quantum regime, but if the system were cooled to $<100$~mK then we estimate that the transfer fidelity for a quantum state would be $\gtrsim10\%$.  Here we first describe a possible quantum state transfer protocol and then estimate the effect of the two dominant mechanisms for decoherence in the system: the thermal Langevin force and the optical force from radiation-pressure shot noise.

\begin{figure}[h!]
\includegraphics[width=.48\textwidth]{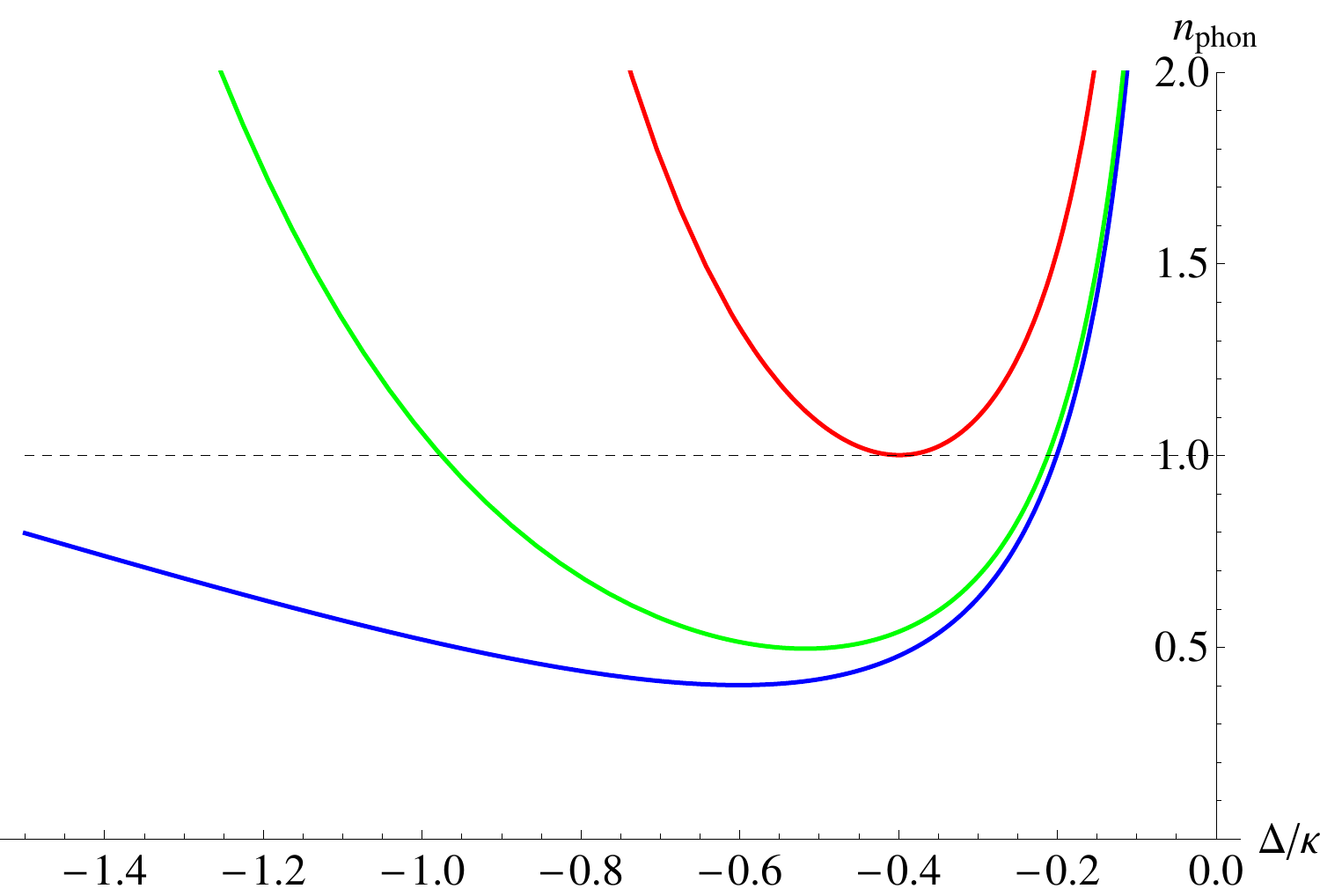}
\caption{Theoretical calculation of the average energy of a mechanical mode (in units of phonons) as a function of the detuning $\Delta$ of the laser from the cavity resonance frequency (in units of the cavity linewidth $\kappa$) for three different incident laser powers: 1.2~\textmu W in red, 10~\textmu W in green, and in the limit of infinite laser power in blue.\cite{WilsonRae2007,Marquardt2007}  We assume a 21~MHz linewidth optical cavity, as in the main paper, and calculate the average phonon occupation of a single 7~MHz mechanical mode in a 100~mK thermal environment with a Q of $5\times10^6$ and a single-photon optomechanical coupling $g/2\pi=850$~Hz.}
\label{fig:n_phonon}
\end{figure}

  One possible quantum state transfer protocol is very similar to the energy transfer experiment described in the main text.  First, the two mechanical modes are optomechanically cooled using a 10-20~\textmu W ``cooling'' laser to close to the ground state, that is, with an average energy that is less than one phonon (see \fig{fig:n_phonon}).  This optical power should be compatible with operation in a dilution refrigerator.  Simultaneously, one of the modes is driven to a low-energy coherent state by, for example, modulating the incident laser power at that mode's mechanical resonance frequency or using a piezo drive.  After the system reaches a steady state, the cooling laser and mechanical drive are turned off and a stronger ``coupling'' laser pulse is applied for a short amount of time.  Transferring the coherent mechanical state between the modes would typically require a 30~\textmu sec pulse with 200~\textmu W of power, which should be compatible with operation at 100~mK. This general approach could also be used to entangle the two mechanical modes by only half-completing the transfer.  Note that, unlike the dissipation-induced steady state entanglement discussed in \cite{Genes2008}, this entanglement relies on the dispersive coupling between the two modes, the scheme is pulse-based, and the state is transient.
  
  It has been pointed out before \cite{Genes2008} that the presence of two nearly degenerate mechanical modes can significantly hinder optomechanical cooling if the frequency difference between the two modes is smaller than the optomechanically-damped linewidths $\delta\gamma_{1,2}$.  One way to think about this effect is that the mechanical modes hybridize into the bright and dark mode basis as the laser power and optomechanical cooling/damping is increased.  Since the dark mode is almost completely decoupled from the cavity field, the cooling and damping of the dark mode is significantly reduced compared to the expectations from single-mode optomechanics.

  If we take the 7~MHz modes described in this paper and assume a 100~mK environment and quality factors of 5 million (consistent with the quality factor of other silicon nitride membranes at cryogenic temperatures \cite{Jayich2012,Purdy2012}), then the cooling-induced decoupling, that is, hybridization, is relatively small.  The optomechanical damping required to cool these modes below the single phonon level is $\sim$2~kHz (close to the thermal decoherence rate discussed below), while the two mechanical modes are spaced by $\sim$5~kHz. However, reductions in the quality factor of either mode or in the frequency difference between the modes could keep the ``darker'' mode from cooling to the single phonon level.
  
  Now we consider the two main sources of decoherence which will limit the quantum transfer efficiency.  The thermal Langevin force results in a decoherence rate for a coherent state that is approximately $S_{FF}^{th} = \gamma_m(2n_{th}+1)$, where $\gamma_m$ is the intrinsic mechanical linewidth, $n_{th}\approx k_BT/(\hbar \omega_m)$ is the thermal phonon occupation, $T$ is the temperature of the environment, and $\omega_m$ is the intrinsic mechanical resonance frequency.  Using the parameters from the previous paragraph, the resulting thermal decoherence rate is approximately 1~kHz which is smaller than our current 30~kHz coupling rate.
  
  The other decoherence channel is due to optical vacuum fluctuations, that is, radiation pressure shot noise (RPSN). The magnitude of the decoherence rate is approximately $S_{FF}^{RPSN} = \delta\gamma(2n_{opt}+1)$. Here $\delta\gamma=-2\mathrm{Im}(\Sigma[\omega_m])$ is the optical damping and $n_{opt}$ is the effective phonon occupation due to the non-zero effective temperature of the optical mode bath. In the resolved-sideband limit ($\omega_m\gg\kappa$) whenever the detuning $\Delta\approx-\omega_m$ then $n_{opt}\ll 1$.  In this regime the effective temperature of the optical bath has negligible effect on the decoherence rate (and the optical damping can cool the mechanical mode to much less than one phonon).  However, in the Doppler regime $\omega_m\lesssim\kappa$ the effective temperature $n_{opt}$ can become more than one which significantly increases the decoherence rate.  Expressing the optomechanical decoherence rate in terms of previously derived system parameters (see the first section of the supplemental material), we find
\begin{eqnarray}
S_{FF}^{RPSN}=|\alpha|^2\kappa(|\chi_c[\omega_m]|^2+|\chi_c[-\omega_m]|^2).
\end{eqnarray}

  Based on these decoherence rates, we can roughly estimate the transfer fidelity of a quantum state.  The optomechanical coupling $\mathrm{Re}(\Sigma_{12})$ between the two mechanical modes sets the approximate transfer time $\pi/\mathrm{Re}(\Sigma_{12})$.  The transfer fidelity is approximately equal to the decoherence during this time
\begin{eqnarray}
\exp \left[-\frac{\pi}{\mathrm{Re}(\Sigma_{12})}(S_{FF}^{th}+S_{FF}^{RPSN}) \right] .
\end{eqnarray}
Since both the decoherence rate $S_{FF}^{RPSN}$ and the coupling $\mathrm{Re}(\Sigma_{12})$ depend on the laser detuning $\Delta$, we maximize this ratio for our system (assuming a thermal decoherence rate of $\sim$1~kHz as given above and 200~\textmu W incident laser power) and estimate a quantum state transfer efficiency of $\sim10\%$.  

  This brief discussion of transfer efficiency also serves to highlight the differences between this experiment and previous optomechanical experiments with coupled mechanical oscillators.  The two membrane modes explored in this experiment have nearly identical frequencies (5~kHz splitting) and large quality factors $Q>100,000$ with linewidths of 15~Hz and 30~Hz.  The frequency difference and linewidths are negligible compared to the maximal coupling rate of about 30~kHz, so the main limitation on transfer efficiency was the optical damping and this limitation can be reduced in the future.
  
  For comparison, earlier experiments have demonstrated larger absolute coupling rates: 300~kHz in \cite{Massel2012}, about 0.5-1~MHz in \cite{Lin2010}, and about 1.5~MHz in \cite{Zhang2012}. However, the 450~kHz frequency splitting between intrinsic modes in \cite{Massel2012} was larger than the coupling rate, thus the intrinsic modes were not fully hybridized which, among other things, results in a poor transfer efficiency. The systems in \cite{Lin2010} can be fully hybridized; however the intrinsic damping rates are comparable to the coupling rates which fundamentally limits the energy transfer efficiency.
  
  The system in \cite{Zhang2012} has a coupling that is much larger than the mechanical frequency splitting of 60~kHz and intrinsic mechanical linewidths of about 20~kHz, thus we would expect efficient transfer of classical energy.  However, we estimate that this linewidth is too large for efficient quantum state transfer.  This experiment also concentrated on a very different regime, in which the optical force was used to drive self-sustained mechanical oscillations and synchronize the two mechanical oscillators.

\clearpage

  \Large{\bf{3.  Cryogenic operation of a fiber-cavity}} \\
\normalsize

  To date, fiber-cavities have been operated at room-temperature.  However, a fiber-cavity is a natural way to integrate an optical cavity into a dilution refrigerator.  This is because it is intrinsically fiber-coupled, which greatly simplifies the task of getting light into and out of the cryostat.  It is also a compact device which easily fits on the cold-plate of a typical refrigerator.  Here we describe the first demonstration of a fiber-cavity (without a membrane) over the temperature range 300~K to 4~K.

\begin{figure}[h!]
\includegraphics[width=.48\textwidth]{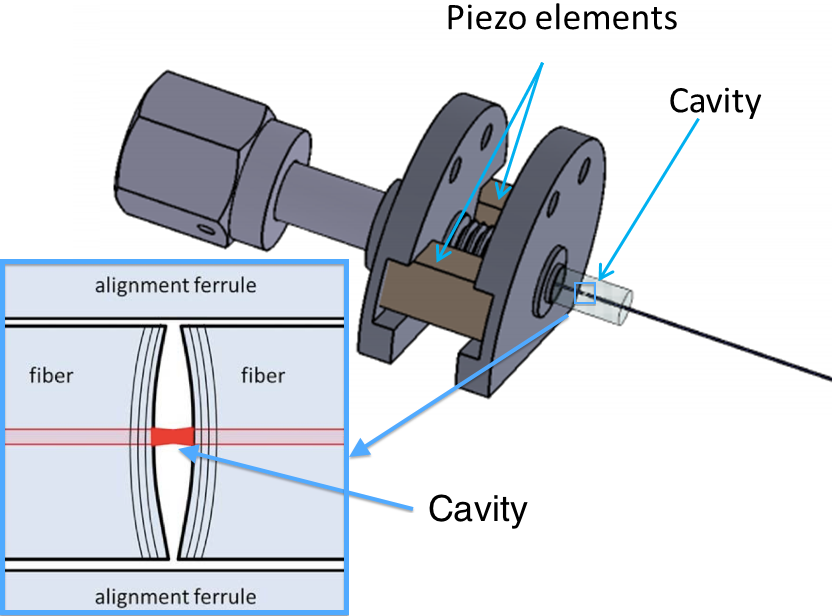}
\caption{Device for cryogenic testing.  The cavity is formed inside of a single glass ferrule and the cavity length is tuned using a pair of piezo elements driven in parallel.  At room temperature the cavity length is ~50 microns.}
\label{fig:cryo_layout}
\end{figure}

  This device (\fig{fig:cryo_layout}) is formed by inserting mirror-coated fibers into each side of a single glass ferrule.  The ferrule aligns the two mirrored fiber-ends along most directions.  After rotating each fiber to optimize finesse, one fiber is glued to the glass ferrule and the other is glued to a metal plate which is actuated by a piezo to change the cavity length.

\begin{figure}[h!]
\includegraphics[width=.48\textwidth]{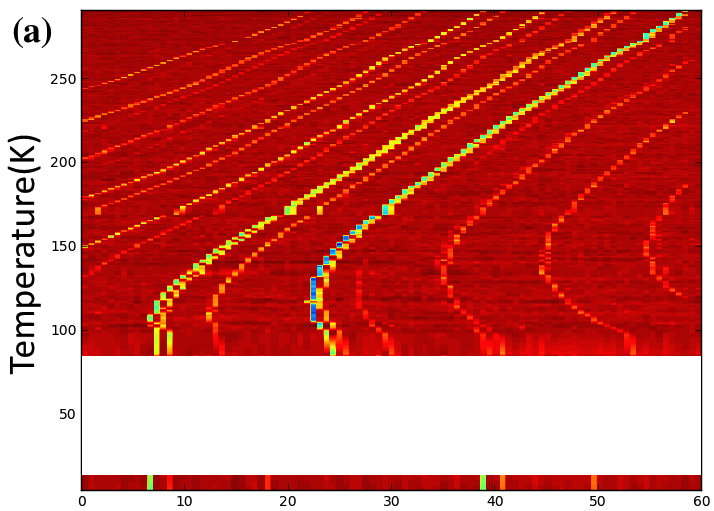}\\
\includegraphics[width=.48\textwidth]{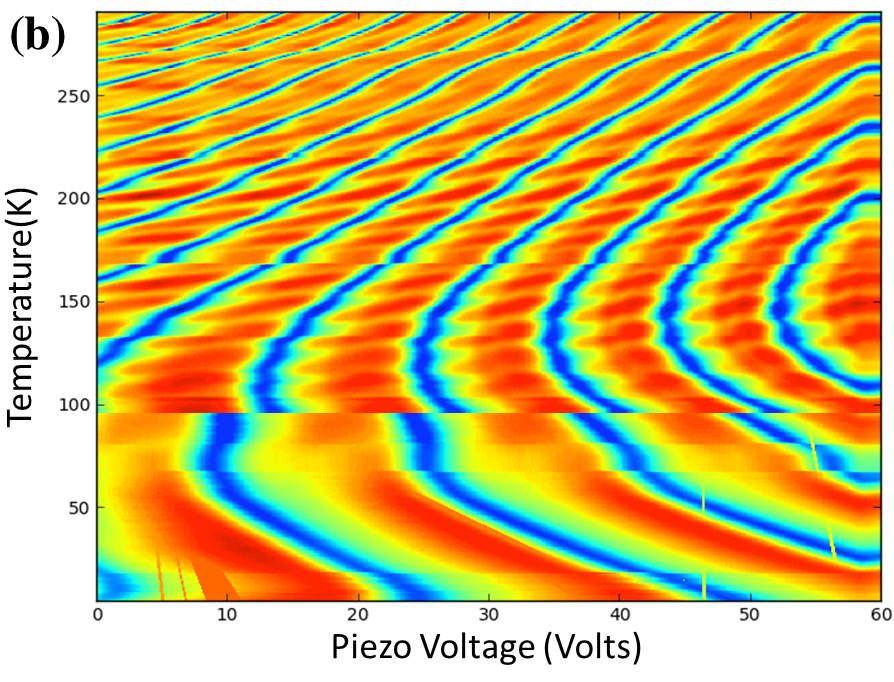}\\
\includegraphics[width=.48\textwidth]{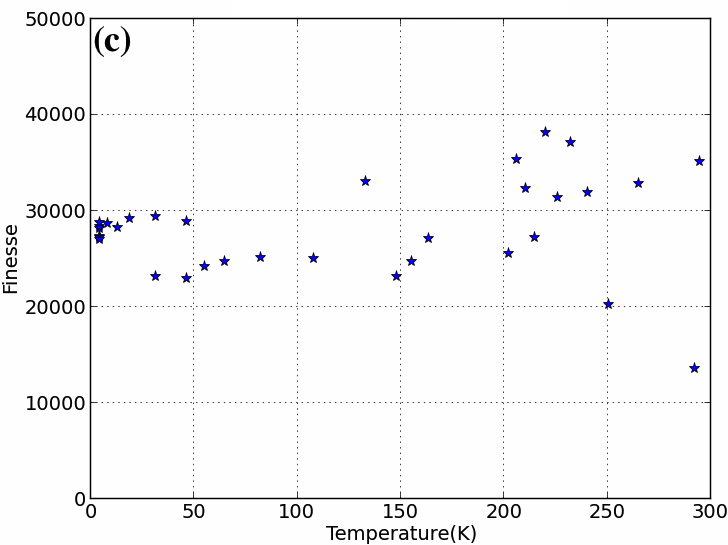}\\
\caption{(a,b) The reflection from the cavity is plotted as a function of voltage applied to the cavity-length piezo (x-axis, larger voltages correspond to a longer cavity) and cavity temperature (y-axis) for two different laser wavelengths:  (a) 1550nm, where the cavity mirrors are highly reflective with finesse of 30,000 (technical difficulties resulted in no data between 80~K and 4~K);  and (b) 1310nm, where the cavity mirrors have a small reflectivity $R\approx10\%$.  As the temperature decreases, we observe that the cavity length decreases (by about 6~\textmu m) and the piezo range also decreases (to about 2~\textmu m).  (c) We also observe that finesse of the cavity at 1550nm is approximately independent of temperature.}
\label{fig:cryo_cooldown}
\end{figure}

  We characterize the performance of the fiber-cavity by measuring the cavity reflection as a function of temperature and piezo voltage at two different wavelengths (1550nm in \fig{fig:cryo_cooldown}(a), 1310nm in \fig{fig:cryo_cooldown}(b)).  The mirror-coating is designed to be highly reflective at 1550nm (finesse of 30,000 at room temperature) but is only weakly reflective at 1310nm ($R\approx10$\%).  As expected, we observe that the cavity length and the piezo travel range are temperature dependent.   However, the cavity finesse (\fig{fig:cryo_cooldown}(c)) and coupling are approximately independent of temperature.
  
  In summary, the cavity is cryogenically compatible because the change in cavity length is small (less than 10~\textmu m), the piezo travel at 4~K is still greater than the laser wavelength, and the cavity finesse and coupling do not change appreciably with temperature.
  
\end{document}